\documentclass{aa}

\usepackage{txfonts}
\usepackage{graphicx}
\usepackage{natbib}
\bibpunct{(}{)}{;}{a}{}{,}
 
\begin{document}
 
\title{A highly collimated, extremely high
velocity outflow in Taurus}

\author{M. Tafalla \inst{1}
\and
J. Santiago \inst{1}
\and D. Johnstone \inst{2}
\and R. Bachiller \inst{1}
}

\institute{Observatorio Astron\'omico Nacional (IGN), 
Alfonso XII 3, E-28014 Madrid,
Spain
\and 
NRC Canada, Herzberg Institute of Astrophysics, 
5071 West Saanich Road, Victoria, B.C. V9E 2E7, Canada
}
 
\offprints{M. Tafalla \email{m.tafalla@oan.es}}
\date{Received -- / Accepted -- }
 
\abstract{
We present the first case of a highly collimated, extremely high velocity
bipolar outflow in Taurus. It is powered by the low-luminosity
(0.4 $L_\odot$) source IRAS 04166+2706 and contains gas accelerated up
to 50 km s$^{-1}$ with respect to the ambient cloud
both toward the blue and the red (uncorrected for projection). 
At the highest velocities, the outflow collimation
factor exceeds 20, and the gas displays a very high degree of spatial 
symmetry. This very fast gas presents multiple maxima, and most likely
arises from the acceleration of ambient material by a time-variable
jet-like stellar wind. When scaled for luminosity, the
outflow parameters of IRAS 04166
are comparable to those of other extremely high
velocity outflows like L1448, indicating that even 
the very quiescent star-formation mode 
of Taurus can produce objects powering very
high energy flows  ($L_{mec}/L_* > 0.15$).
}

\authorrunning{Tafalla et al.}
\titlerunning{An extremely high velocity outflow in Taurus}
 
\maketitle

\section{Introduction}

The ejection of supersonic gas in a bipolar outflow is one of the 
first signatures of stellar birth and very likely a required element
in the physics of star formation. Outflows have been found
towards young stellar objects (YSOs) of nearly all
masses, and in most cases, their molecular
component (traced in CO) presents a weak degree of collimation
and a prevalence of low velocity gas \citep{lad85}.
A small group of outflows, however, appears dominated by 
an extremely high velocity (EHV) molecular component having a speed over 
20 km s$^{-1}$ (Mach $> 100$ for gas at 10 K) and a very high degree of
collimation (length/width $> 10$, see \citealt{bac96}). This group of highly 
collimated EHV outflows seems associated with the youngest protostars 
(Class 0, \citealt{and93}), and probably represents a specially energetic
initial stage in the evolution of a bipolar flow (e.g., \citealt{bon96}).

Although the nearby Taurus cloud contains
a large population of outflows \citep{hey87,mor92,and99},
no Taurus YSO has been found so far
to power an EHV flow.
This is surprising since other nearby
clouds like Perseus and Ophiuchus contain 
EHV flows \citep{bac90,bac91,mcc94,and90},
and it may at first suggest that 
the gas conditions in Taurus prevent the formation of this type
of systems.  In this paper, however, we report the identification 
of the first highly collimated EHV outflow in Taurus, which is 
associated with 
the low luminosity (0.4 L$_\odot$) source IRAS 04166+2706 (I04166
hereafter). The youth of this source has been previously recognized  
\citep{hey87,ken90,bar92,che95,shi00,mot01,par02},
but only a poorly collimated
outflow and weak [SII] emission have been reported so far 
\citep{bon96,gom97}. In the following 
sections we present
the newly found EHV outflow
and discuss some implications for 
outflow and star-formation studies. A future paper will describe the 
shock-induced chemistry of this flow (Santiago et al. 2004, in preparation).

\section{Observations}

We observed CO(J=2--1) with the 9-pixel HERA receiver array
on the IRAM 30m telescope in 2004 April. In order to obtain flat 
baselines, we observed in wobbler switching (WSW) mode chopping 
$240''$ away in azimuth. This procedure 
contaminated the spectra at low velocities due to the extended Taurus 
emission, so we also observed in position
switching (PSW) mode using a reference more than 1 degree away
and free of CO emission at the velocities of interest; these observations
have been used to reconstruct the emission at low velocities.
In addition, 4 selected positions were observed in CO(1--0)
using WSW mode to help estimating excitation conditions.
All these data were taken with the VESPA correlator set to 
velocity resolutions of 0.4 and 0.8 km s$^{-1}$ for CO(2--1) and
CO(1--0), respectively. Cross scans of bright continuum sources
were used to correct the telescope pointing, and standard 
efficiencies were used to convert the data into the main beam brightness
scale. The FWHM of the telescope beam was $22''$ and $11''$ for
CO(1--0) and CO(2--1), respectively.

\begin{figure*}[t]
\centering
\resizebox{15cm}{!}{\includegraphics{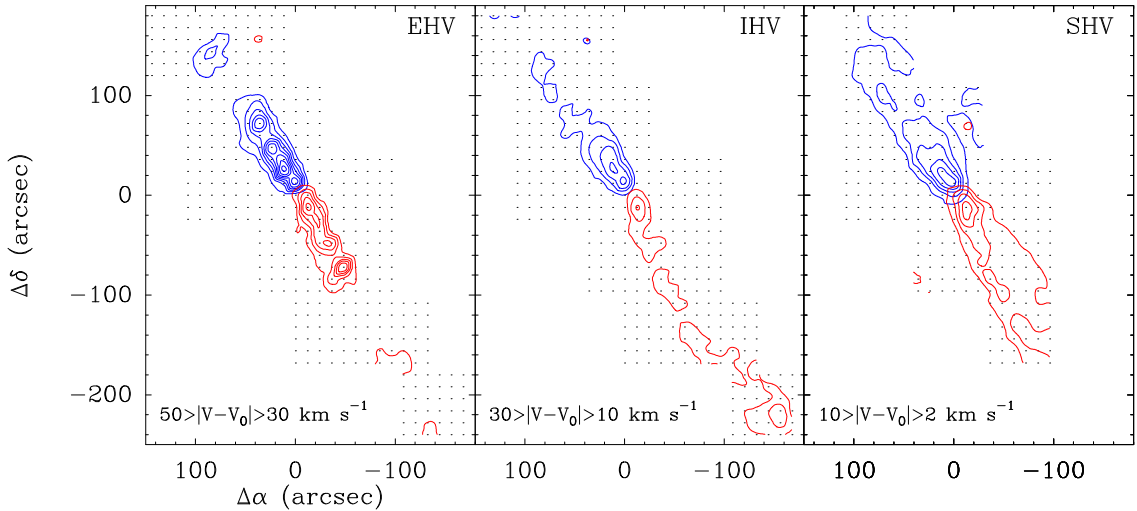}}
\caption{CO(2--1) integrated intensity maps for the EHV, IHV, and
SHV regimes (see text). Note the high symmetry between the EHV blueshifted
gas (blue contours to the NE) and the EHV redshifted gas (red contours to 
the SW). The EHV and IHV maps have been made using WSW spectra, 
while the SHV map has been made using PSW data. 
First contour and interval are 1 K km s$^{-1}$ in all plots
except for the more prominent blue
gas in the rightmost (SHV) panel, where they have been increased 
to 3 K km s$^{-1}$ to avoid overcrowding.
Offsets are referred to the nominal position of IRAS 04166+2706
($\alpha_{J2000}=4^{\rm h}19^{\rm m}42\fs6$, 
$\delta_{J2000}=27\degr13'38''$), and $V_0=6.7$ km s$^{-1}$ has been 
derived from NH$_3$(1,1) data.
\label{fig1}}
\end{figure*}

We observed NH$_3$(1,1) and NH$_3$(2,2) simultaneously toward I04166
with the 100m telescope of the MPIR at Effelsberg in 1998 October. We 
used frequency switching mode and the AK90 autocorrelator to 
achieve a velocity resolution of 0.03 km s$^{-1}$. Cross scans
of continuum sources were used to correct the telescope pointing, and 
an observation of L1551 was used to calibrate the data. 
The telescope beam FWHM was $40''$.

\section{The outflow}

\subsection{Morphology and kinematics}

Figure 1 presents integrated maps of the CO(2--1) emission toward I04166
for three velocity ranges. These ranges are offset on average from the cloud 
systemic velocity ($=6.7$ km s$^{-1}$) by 40, 20, and 8 km s$^{-1}$,
and will be referred hereafter as the extremely high velocity (EHV), 
intermediate
high velocity (IHV), and standard high velocity (SHV) regimes, respectively.
In all maps, the CO emission forms two well-separated lobes
that have I04166 near their apex, clearly indicating 
the presence of a bipolar outflow powered by this source. 
At the most extreme velocities, the outflow emission 
is unresolved transversally by our $11''$ beam, so even if we only 
consider the innermost (connected) region of emission ($220''$ long), 
the implied collimation factor is at least 20. This factor
makes the I04166 system one of the most collimated outflows known 
(see \citealt{bac96}).

As Figure 1 shows, each EHV component consists of a
connected lobe $110''$ (0.07 pc) long near I04166
and additional emission at further distance. Both
blue and red connected lobes break up into peaks almost
symmetrically located with respect to the IRAS source.
Each of these lobes ends in a prominent peak, and
while the blue lobe contains three additional maxima,
the red one has two and a hint of a third one in between. 
These presence of symmetric peaks is common in EHV outflows
from low mass stars, like L1448 and I03282 (so-called ``bullets'':
\citealt{bac90,bac91}), and is suggestive of a 
discontinuous, even periodic, mass ejection mechanism.
Dividing the average
CO speed of the EHV gas (40 km s$^{-1}$) by the average peak separation
($20''$), we infer an ejection period of 350 yr.

More than $110''$ away from the IRAS source, both EHV lobes
contain additional discrete peaks. The prominent blue peak near
($85''$, $145''$) seems to have a counterpart near ($-100''$, $-170''$)
(unfortunately not fully mapped due to lack of time), and there is a hint
of an additional EHV peak in the southernmost part of the blue lobe. 
This peak 
is probably part of a bright IHV component, and may have a weak blue
counterpart at the northernmost part of our map. Overall, the EHV emission
shows a high degree of blue-red symmetry.

The IHV emission (middle panel of Fig. 1) shows a more diffuse distribution 
and a lower degree of symmetry (but see energetics below). 
The blue IHV lobe follows the EHV emission and deviates
slightly to the east at large distance from the IRAS source. The red lobe
also deviates east at intermediate distances but lies again close to
the outflow axis by the end of our map, where it presents a discrete
peak. As mentioned before, there is a hint of blue IHV emission 
at the north end of our map, but even if we discount this region
(which needs further mapping), the IHV extends more than $450''$,
which corresponds to more than 0.3 pc at the distance of Taurus (140 pc).
Dividing this projected distance by the average radial 
velocity extent of this component
we derive a dynamical age of 7500 yr, and using the highest outflow
speed reduces the estimate to 3000 yr.

Finally, the SHV emission forms two conical lobes with
the same axis as the EHV gas and
relative maxima close to the apex.
Again there is a slight enhancement toward
the east,  but the bulk of the SHV emission is rather
symmetric and envelopes the EHV emission. This suggests 
a stratification of the outflow gas, with the
EHV gas moving along the axis and the slower SHV material at further
distances (see \citealt{gue99} for a similar case in HH211). The
overall high collimation of the outflow allows to separate it from 
the [SII] jet-like feature HH 390 identified by 
\citet{gom97}, which seems powered by a separate source almost $2.'5$
north of I04166 (J. Bally and J. Walawender, private communication).

\subsection{Energetics and central core}

\begin{figure}
\centering
\resizebox{8cm}{!}{\includegraphics{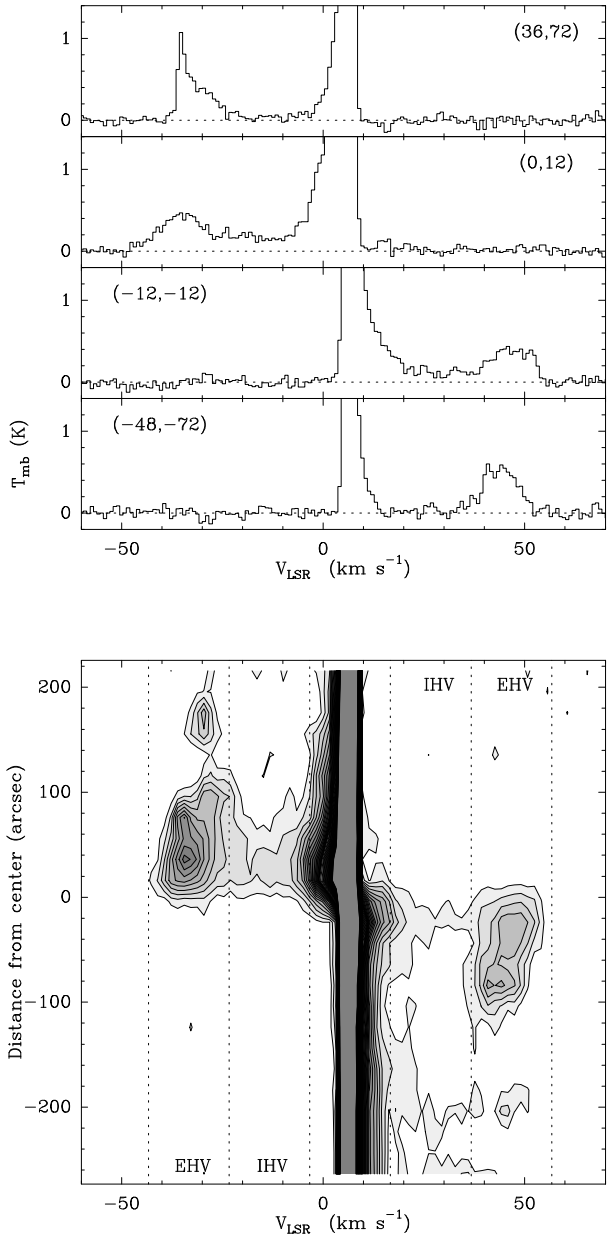}}
\caption{{\bf Top:} CO(2--1) spectra toward selected outflow positions.
Note how the EHV gas appears as a distinct feature in the
spectra. {\bf Bottom:} Position-velocity diagram of the CO(2--1)
emission along the full length of the outflow, also showing
the distinct nature (and near constant velocity) of the EHV component.
First contour and interval are
0.05 K. All data are from WSW observations with the
emission between $V_{lsr}$ -3.3 and 16.7 km s$^{-1}$
reconstructed from PSW observations.
Offset origin as in Fig. 1.
\label{fig2}}
\end{figure}

Fig. 2 shows a series of CO(2--1) spectra
(top) and a position-velocity (PV) diagram along the outflow (bottom)
illustrating the three velocity regimes. 
The slower SHV gas appears as a wing in the spectra, 
while the weaker IHV gas forms a flat, low
level component more extended toward the south. 
The EHV gas, on the other hand, forms discrete maxima in the spectra, 
like the EHV features found in other highly 
collimated outflows \citep{bac90,bac91}.
This component seems to decrease in velocity slightly 
along the outflow axis ($<10$ km s$^{-1}$), and 
its spectral signature varies 
significantly with position (compare the top two spectra).

Given the distinct nature of the velocity components (see Sect. 4 for 
further discussion), we estimate their energetics content separately.
We assume that the CO(2--1) emission is optically 
thin and we do not correct for projection, so our results are strict 
lower limits. We estimate the CO excitation 
temperature from the 2--1/1--0 ratio at
selected outflow positions, which we find 
to range 
between 1 and 2.5. Since these ratios imply 
excitation temperatures between 7 and 20 K, 
we use a mean value of 15 K, as also
suggested by the highest S/N CO(1--0) spectra, and which gives rise 
to the lowest column density for a given integrated intensity.
We also assume a CO abundance of $8.5 \times 10^{-5}$
\citep{fre82}.

\begin{table}
\caption[]{Outflow energetics
\label{tbl-1}}
\[
\begin{array}{lccccccc}
\hline
\mbox{} & \multicolumn{2}{c}{\mbox{SHV~~}} &
\multicolumn{2}{c}{\mbox{IHV~~}} &
\multicolumn{2}{c}{\mbox{EHV~~}} & {\mbox{TOT}} \\
\noalign{\smallskip}
\hline
\noalign{\smallskip}
\mbox{M~~ ($10^{-3} M_\odot$)} & 7.7 & (3.1)^a\;\;\; & 1.1 & (1.0)^a\;\;\;
& 1.3 & (1.2)^a\;\;\; & 10 \\
\mbox{P~~ ($10^{-2} M_\odot$ km s$^{-1}$)~~~} & 2.9 & (2.7)^a\;\;\; &
2.0 & (1.0)^a\;\;\; & 4.9 & (1.2)^a\;\;\; & 10 \\
\mbox{E~~ ($10^{42}$ erg)} & 2.6 & (2.5)^a\;\;\; & 8.2 & (1.1)^a\;\;\; &
38 & (1.2)^a\;\;\; & 50 \\
\noalign{\smallskip}
\hline
\end{array}
\]
\begin{list}{}{}
\item[$^{\mathrm{a}}$] Values in parenthesis are blue/red ratios
\end{list}
\end{table}

The results of our outflow energetics are shown in Table 1, and 
can be summarized by saying that the outflow has a mass of at least
0.01 $M_\odot$, a momentum of 0.1 $M_\odot$ km s$^{-1}$, and
an energy of $5 \times 10^{43}$ erg. A look at the distribution of 
these parameters among the outflow components shows that most of
the mass lies at low velocities while most of the momentum and
energy is in the fastest gas. Quantitatively, the EHV component 
contains only 13\% of the mass but carries 50\% of the outflow
momentum and more than 75\% of its kinetic energy. 

Comparing the parameters of the blue and red gas
(ratios in parenthesis in Table 1),
we find a high degree of symmetry 
in the outflow energetics: the blue/red ratio 
is 1.2 for the EHV gas and 
even closer to 1.0 for the IHV component, despite the
lower symmetry of its CO spatial distribution (Fig. 1).
The SHV regime, on the other hand, has a factor of 3 more 
blue gas, but the study of this component is limited
by the ambient emission. The energetics measurements, therefore, 
confirm the indications of the maps in Fig. 1 that at least the
fastest part of the I04166 outflow has a very high degree of symmetry.

To complete the description of the I04166 system we present in Fig. 3 
NH$_3$(1,1) and 1.2mm continuum maps of the IRAS source vicinity
(see \citealt{mot01} for a similar continuum map). These
maps show that the dense core  
is nearly round and clearly peaked toward the IRAS source, and that it
forms part of a SE-NW chain of cores (to be discussed in a future paper).
Using standard NH$_3$ analysis, we derive a gas kinetic
temperature of 10 K, a turbulent component of 
0.2 km s$^{-1}$ (FWHM),
and assuming a typical Taurus average 
para-NH$_3$ abundance of $5 \times 10^{-9}$ \citep{taf04}, we
derive a mass of 1.5 M$_\odot$, all typical of a Taurus core 
\citep{jij99}.

\begin{figure}
\centering
\resizebox{\hsize}{!}{\includegraphics{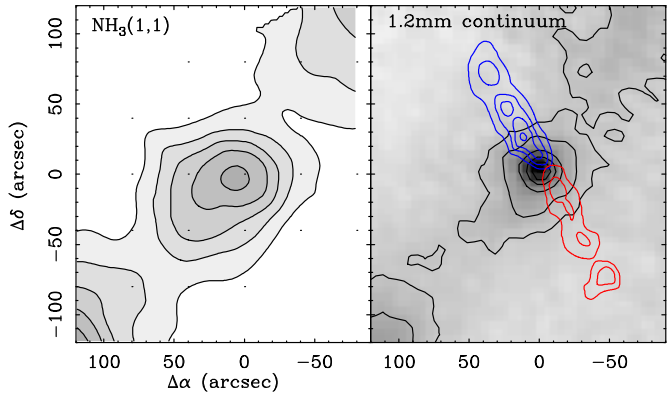}}
\caption{{\bf Left:} Integrated intensity map of the
NH$_3$(1,1) emission toward the I04166 dense core.
First contour and interval are 2 K km s$^{-1}$
{\bf Right:} Map of 1.2mm continuum emission toward
the I04166 core (from Tafalla et al., in preparation)
with several contours of the EHV outflow map from Fig. 1.
The continuum emission has been convolved with a $10''$
beam to enhance sensitivity, and its first contour
and interval are 20 mJy per $11''$-beam.
Offset origin as in Fig. 1.
\label{fig3}}
\end{figure}

\section{Discussion}

The most striking element of the I04166 outflow is the EHV gas, with its
high degree of collimation, large radial 
velocity, and strong blue/red symmetry. Because of its 
limited location and
high velocity dispersion ($>10$ km s$^{-1}$, Fig. 2), the EHV gas
is unlikely to represent the (still mysterious) agent that drives
the outflow, although its extreme properties suggest 
that it must be closely connected to the outflow driving agent. 
Given the geometry of the EHV gas (Fig. 1),  
this agent has to be jet-like 
(or have an important jet-like component), and in this context it 
seems natural to interpret the EHV gas as ambient cloud material 
prompt-entrained by a jet bow shock.
Numerical simulations of jet-driven outflows (e.g., \citealt{lee01}),
however, predict
that gas accelerated by a bow shock will present a PV diagram with
a ``spur'' feature having the fastest gas at the tip of
the jet, in contrast with the almost constant velocity of the separate
EHV component seen in Fig. 2. A possible solution to this discrepancy 
is the existence of multiple bow shocks, as suggested by the
multiplicity of peaks in the maps of Fig. 1. High resolution 
observations of the I04166 outflow and realistic models of 
gas acceleration by time/angle
variable jets are needed to test this interpretation and
attempt to understand the origin of the EHV gas, a problem not peculiar
to the I04166 system but common to all EHV outflows 
(e.g., \citealt{bac96}). 

Comparing the I04166 outflow to other EHV outflows like L1448, 
I03282, or VLA1623 we conclude that it
is a bone fide member of the class 
for its morphology, kinematics, and energetics content.
The I04166 outflow, in fact, seems a scaled-down version of the L1448 
outflow, as it contains about 5\% of its momentum and energy while
it has a central source with about 5\% of its luminosity. 
This simple scaling
(plus a similar kinematic age) shows that the I04166 outflow is almost as
mechanically efficient ($L_{mec}/L_* > 0.15$) as L1448, 
and therefore belongs to the 
group of outflows an
order of magnitude more powerful than typical 
Class I outflows \citep{bon96}. The I04166 source, on
the other hand, is a 
typical Taurus object both in the parameters of its dust
envelope \citep{mot01} as well as in the turbulence and
temperature of its core (Sect. 3.2).
This implies that the (still unexplained) conditions required
to produce an EHV outflow can be realized in the quiescent
Taurus environment, and in addition, they can be achieved
by a source with a luminosity as low as 0.4 L$_\odot$ (the lowest
of all EHV sources). Such a combination of circumstances argues for
EHV outflows resulting from a special (early) evolutionary stage in
the protostellar phase, almost independent of their luminosity and
environment.

Finally, we note that despite the similarities between the I04166
outflow with the outflows from L1448, I03282, or VLA1623,
the spectral energy distribution of I04166 
is characterized by a bolometric temperature $(T_{bol})$ of 90-140 K,
while the $T_{bol}$ from the other sources with EHV outflows 
is closer
to 30-60 K \citep{che95,shi00}. This 
relatively large $T_{bol}$ value makes I04166 a 
Class I candidate \citep{che95}, and therefore 
unexpectedly ``evolved'' compared with the other exciting sources
of EHV flows. The recent non detection of a K band point source
associated with I04166 by \citet{par02}, however, suggests that 
I04166 may still be a truly Class 0 source. 
A better characterization of I04166 in the FIR 
with current (Spitzer) and future (Herschel) telescopes 
is therefore necessary.

\acknowledgements We thank the staff of the IRAM 30m telescope 
for help during the observations, and J. Bally and J. Walawender
for information on the HH390 system. MT, JS, and RB acknowledge support 
from grant AYA2003-07584 of the Spanish DGES.

\bibliographystyle{apj}

\end{document}